\documentclass[prd, amsfonts, preprint, nofotinbib, showpacs, superscriptaddress]{revtex4}

\usepackage{graphicx, epsfig}
\usepackage{color}
\usepackage{enumerate}

\newcommand{\be}{\begin{equation}}
\newcommand{\ee}{\end{equation}}
\newcommand{\bea}{\begin{eqnarray}}
\newcommand{\eea}{\end{eqnarray}}

\newcommand{\gapp}{\mathrel{\raise.3ex\hbox{$>$}\mkern-14mu \lower0.6ex\hbox{$\sim$}}}
\newcommand{\lapp}{\mathrel{\raise.3ex\hbox{$<$}\mkern-14mu \lower0.6ex\hbox{$\sim$}}}
\def\bbox{{\,\lower0.9pt\vbox{\hrule \hbox{\vrule height 0.2 cm
\hskip 0.2 cm \vrule  height 0.2 cm}\hrule}\,}}
\newcommand{\beq}{\begin{equation}}
\newcommand{\eeq}{\end{equation}}

\newcommand{\Mbh}{M_{\rm BH}}
\newcommand{\Mpl}{M_{\rm Pl}}

\begin{document}
\title{Could any black holes be produced at the LHC?}

\author{Jonas Mureika}
\thanks{jmureika@lmu.edu}
\affiliation{Department of Physics, Loyola Marymount University, 
Los Angeles, CA}

\author{Piero Nicolini}
\thanks{nicolini@fias.uni-frankfurt.de}
\affiliation{Frankfurt Institute for Advandced Studies (FIAS) 
and Institut f\"ur Theoretische Physik, Johann Wolfgang Goethe-Universit\"at, Frankfurt 
am Main, Germany}

\author{Euro Spallucci}
\thanks{spallucci@ts.infn.it} 
\affiliation{Dipartimento di Fisica, Universit\`a di Trieste and INFN, Sezione di Trieste, Trieste, Italy
}


\begin{abstract}
We introduce analytical quantum gravity modifications of the production cross section for
terascale black holes by employing an effective ultraviolet cut off $l$. 
We find the new cross sections approach the usual ``black disk'' form at high energy,
while they differ significantly near the fundamental scale from the
standard  increase with respect to $s$.  We show that 
the heretofore discontinuous step function used to represent
the cross section threshold can realistically be modeled by two functions
representing the incoming and final parton states in a high energy collision.
The growth of the cross section with collision energy is thus a unique
signature of $l$ and number of spatial dimensions $d$.
Contrary to the classical black disk result, our cross section
is able to explain why black holes might not be observable in LHC
experiments while they could be still at the reach of ultra-high energy cosmic ray events.

\end{abstract}


\maketitle

\section{Introduction.}
Despite decades of theoretical advances and experimental progress, we are in 
reality no closer
to a formulating a workable framework of quantum gravity -- let alone finding
related definitive experimental evidence  --  and must be content
to speculate about its nature and phenomenology. 
This is not be only an academic exercise: probing a deeper understanding
of the physical processes potentially able to unveil quantum gravity 
signatures would be a major breakthrough in an otherwise stagnating and 
discouraging situation.  In this spirit, we return to the foundational
connection between classical and quantum gravitation, {\it i.e.} Hawking radiation.
Thanks to its robustness in the semi-classical limit, 
Hawking radiation is a widely accepted  benchmark for any reliable theory 
of quantum gravity.

Unfortunately, the chances of a direct detection of the Hawking radiation are 
remote. Since $T \sim M_{\rm BH}^{-1}$, astrophysical 
black holes are too big to display any relevant 
quantum mechanical effects.  It has alternatively been conjectured that smaller
primordial black holes with masses $M_\mathrm{BH}\sim 10^{11}\ \mathrm{kg}$ and 
radii $r_H\sim 10^{-16} \ \mathrm{m}$ may have formed in the extreme density 
flucations of the early universe.  With temperatures 
$T\sim 10^{12}\ \mathrm{K}$, these black holes would be so bright that we 
should be able to observe them, but as of now the Fermi Gamma-ray Space 
Telescope satellite has been unsuccessful in detecting any such evidence 
\cite{glast}.
At even shorter length scales, one enters the domain of
 modern particle physics accelerators.  It has been suggested, however,
 the possibility of a ``particle black hole'' is very unlikely: the energy densities
 required to squeeze a mass completely inside its own gravitational radius is of 
 the order of the Planck mass $\Mpl$ \cite{thorne}, almost $15$ orders of magnitude higher than the 
 LHC energy and eight orders higher than the most energetic cosmic ray ever detected 
 \cite{cr}. It may thus seem Hawking 
 radiation, and maybe also any hint quantum gravity, is inaccessible at least 
 in the immediate future.  

The advent of large extra spatial 
 dimensions accessible at a fundamental scale to $M_\ast\sim 1\ \mathrm{TeV}$
 allows such gravitational 
 collapse to occur for matter compressed at distances of the order of 
 $10^{-4}$ fm \cite{Dimopoulos}.  Despite the myriad fascinating possibilities unveiled in such scenarios, 
the theoretical foundations are far from being understood.  
Perhaps most problematic is the major limitation concerning the description
of micro-black holes, in that it is impossible to correctly describe
end-stage black hole evaporation in the semiclassical limit when $T\sim \Mbh\sim M_\ast$.

%
%

Utilizing the current literature base of inadequate classical metrics, one cannot take into 
account the local loss of resolution which plagues the spacetime when it is probed at high 
energies/short scales. We thus propose a new framework to describe in an effective way 
the nature of a quantum spacetime 
and its signatures  in the physics of microscopic black holes.
As a preliminary step we seek to address 
the fundamental question: how does quantum gravity affect microscopic black hole formation? 

\section{Black hole production.}
The standard expression for the semi-classical black hole production cross 
section is a translation of the ``hoop conjecture'' \cite{thorne} (for improved versions of this result and comments see \cite{EG,Hsu,Cheung:2002aq,Meade:2007sz,Park:2011je})
a black hole is produced whenever
a parton of energy $\sqrt{s}$ hits a target with
 an impact parameter $b < r_H$, {\it i.e.} smaller than the
Schwarzschild radius of the effective two-body
 system,
 \begin{equation}
 \frac{1}{2\pi\, b}\frac{d\sigma\left(\, s\ ; b\,\right)}{db}=
 \Theta_H\left(\, r_H\left(\, s\,\right) - b\,\right)~~,
 \end{equation}
Here, $\Theta_H$ is the Heaviside step-function implemeting
 the constraint $b \le r_H=2G_N\, \sqrt{s} $. 
 Since the impact parameter $b$ is not observable, it must
 be integrated over to obtain the experimentally
 measurable production cross section:
 \begin{equation}
 \sigma\left(\, s\, \right)=2\pi \int_0^\infty db\, b\,
 \Theta_H\left(\, r_H\left(\, s\,\right) - b\,\right)=
 \pi\, r_H^2\left(\, s\,\right)
 \label{bd}
 \end{equation}
We thus recover the ``black disk'' cross section typically
found in the literature.

 For a neutral, non-spinning, black hole of mass $\Mbh=\sqrt{s}$ in
 $d+1$ dimensions, the horizon radius is
 \begin{equation}
 r_H=\left(\, 2G_\ast\,\right)^{1/(d-2)}\, s^{1/2(d-2)}~~,
 \label{rhd}
 \end{equation}
 where we have introduced the higher dimensional gravitational
 coupling constant $G_\ast\equiv l_\ast^{d-1}=M_\ast^{1-d}$.
We note a peculiar feature of black hole ``particle'' physics. 
As mass (energy)
increases, the linear dimension of the black hole {\it increases}, 
in contrast to the expected behavior of normal particles
whose effective scales are determined by the Compton wavelength. This unique behavior
leads to a possible UV self-completion of quantum gravity \cite{SA11}.
 By inserting equation (\ref{rhd}) in (\ref{bd}), we get
 \begin{equation}
 \sigma\left(\, s\, \right)=\pi\, 
 \left(\, 2G_\ast\, \sqrt{s}\,\right)^{2/(d-2)}.
  \label{badsigma}
  \end{equation}
 For typical LHC energies $\sqrt{s}\sim 1-10$ TeV, one obtains cross sections of the order
$\sigma\sim 1$ nb. Given the most recently reported LHC peak luminosity 
$L\sim 3.65\times 10^{37}\ \mathrm{m}^{-2} \ \mathrm{s}^{-1}$ \cite{cern}, 
this would imply that about ten black hole per second would form.
In hindsight we find that, according to the black disk cross section, black holes would have formed at a non-negligible rate even in early particle physics experiments.  By combining (\ref{badsigma}) with Super Proton Synchrotron (SPS) parameters $\sqrt{s}\sim 630 \ \mathrm{GeV}$ and $L\sim 3.6\times 10^{33}\ \mathrm{m}^2 \ \mathrm{s}^{-1}$, one finds that roughly one black hole a day would have formed in 1985 \cite{DeRaad:1985ye}. 

\begin{figure}
 \begin{center}
\includegraphics[height=10cm]{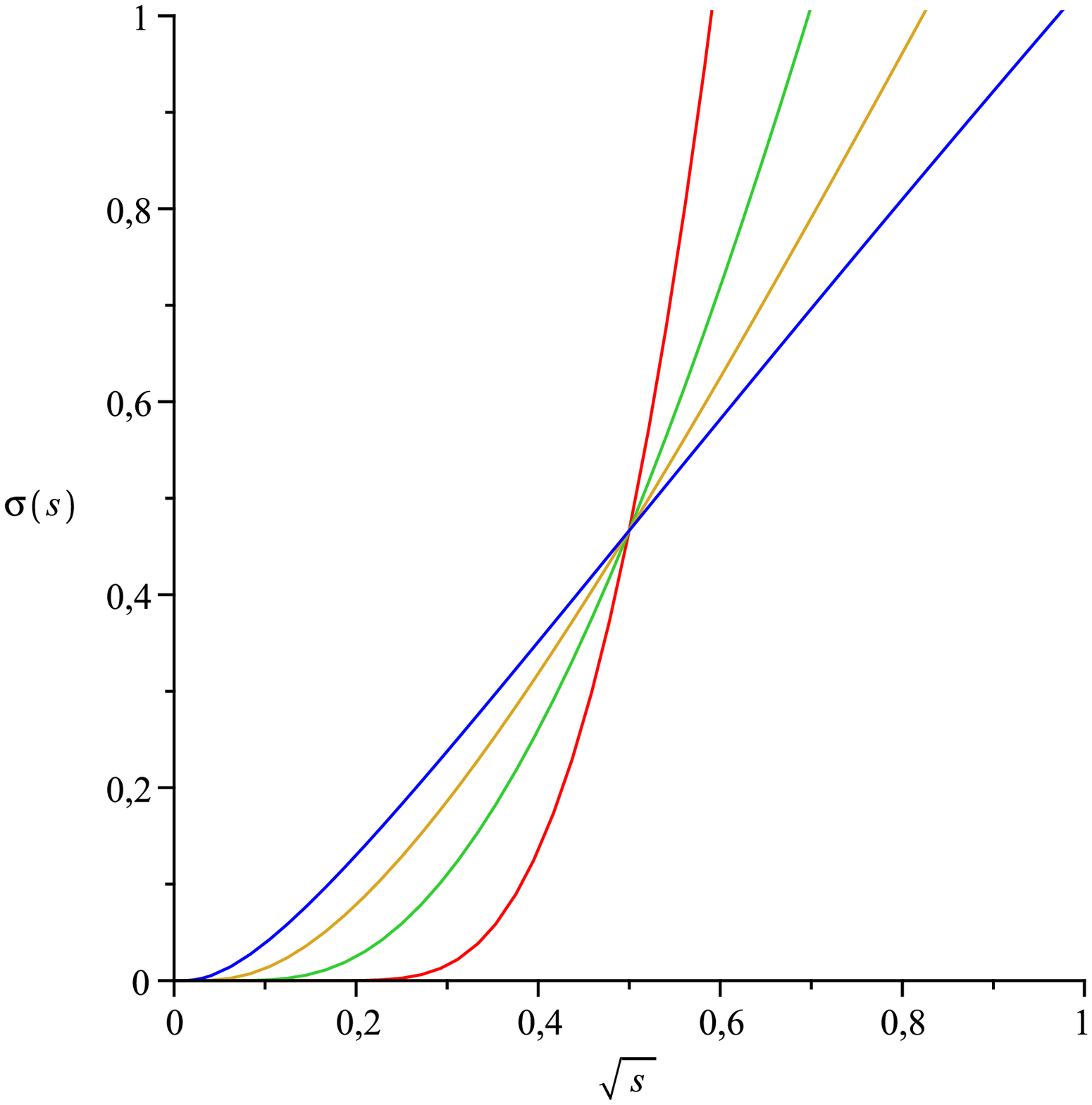}
\includegraphics[height=10cm]{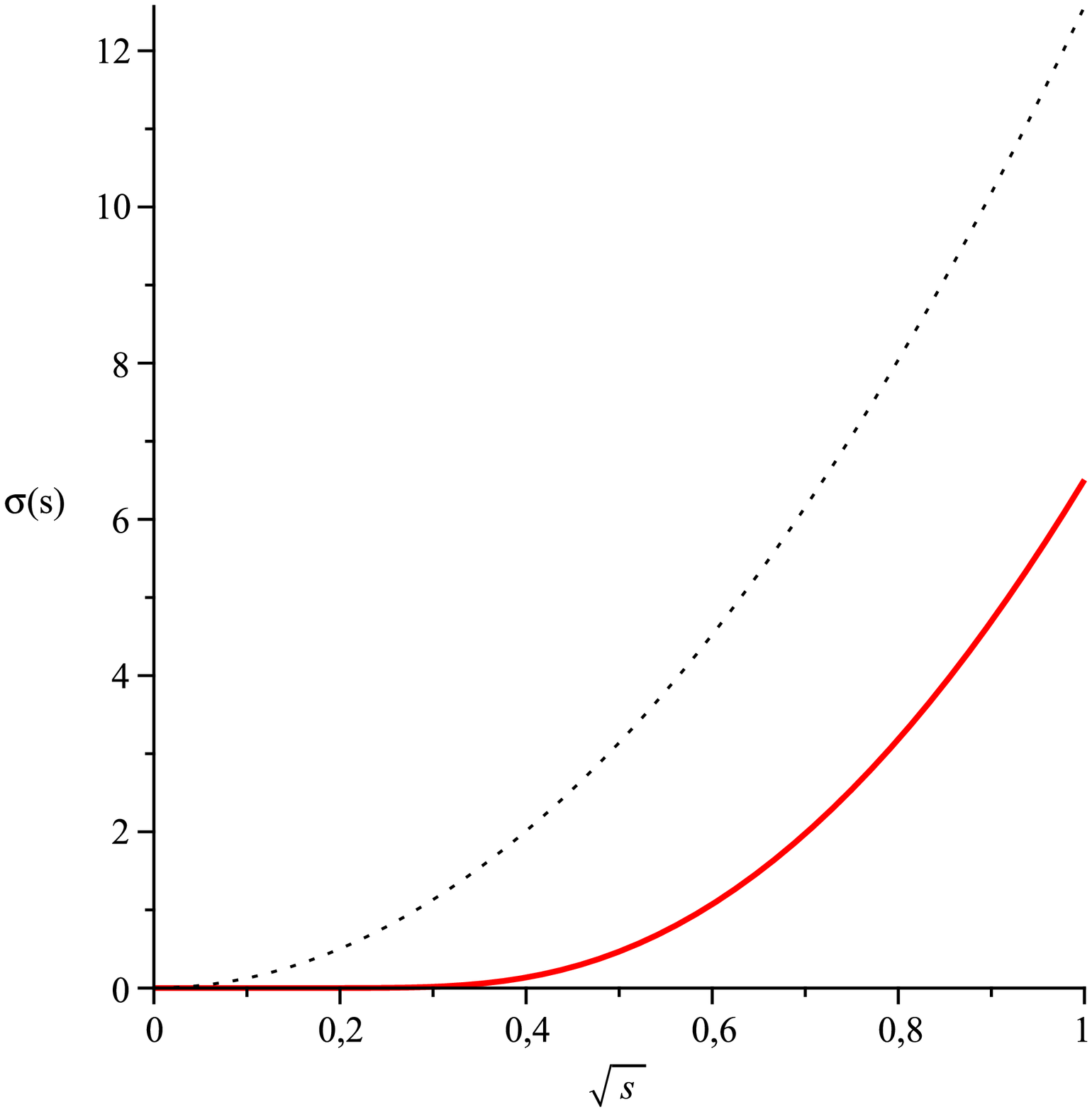}

       \caption{\label{fig1} {Top: Black hole cross sections (\ref{crsect}) as a function of $\sqrt{s}$ for different values of $d$ in $l_\ast$-units (from top to bottom on the right: $d=3, 4,5, 6$). All values are modified at low energies while matching the standard predictions at large $s$. Bottom: The same plot as above for $d=3$ (solid curve) with different ordinate scale to facilitate comparison with the the standard black disk approximation (dotted curve) in the low-energy regime $\sqrt{s}<l_\ast^{-1}$.}}
     \end{center}
  \end{figure}

The aforementioned production rate estimates have since been improved to much lower values \cite{Cheung:2002aq,Meade:2007sz,Park:2011je}. These results, however, conflict with the latest experimental investigations that effectively rule out  the possibility of black hole formation at the LHC \cite{CMS}, at least as far as the semiclassical regime is valid \cite{CHG}. 
The weak point of the hoop conjecture is that for any $s$ black holes
 can be produced provided $b$ is small enough.  On the contrary,
we expect the black hole production channel to open
 only above some threshold energy.  The inaccuracy of predictions due to (\ref{badsigma})  
 follows from
 the assumption that the impact parameter, $b$, can take on aribitrary
 small values, which is not the case in any theory of quantum gravity where
 a minimal length emerges as a new fundamental constant of nature \cite{hep-th/0405033}.
We thus have to introduce a non-vanishing lower integration limit into
 equation (\ref{bd}) to account for the breakdown of any semi-classical
  description of spacetime in a true quantum regime. 
  In theories with large extra-dimensions, the constraint 
 $l_\ast >> l_{Pl.} $ yields
 quantum gravitational excitations. 
Any chance to observe at least
 some indirect signal of quantum gravitational phenomena at LHC requires
 taking $l_\ast$ as the ``minimal length''.   Moreover, instead of demanding a
 cut-off $b\ge l_\ast$ in the scale size, we introduce
 a proper exponential suppression that drives the integral to zero 
 faster than any power of $s$,
  \begin{equation}
 \sigma\left(\, s\, \right)=2\pi \int_0^\infty db\, b\, e^{-l_\ast^2/b^2}
 \Theta_H\left(\, r_H\left(\, s\,\right) - b\,\right).
 \label{good}
 \end{equation}
 Integration can be carried out and gives 
 \begin{equation}
 \sigma\left(\, s\, \right)=\pi\, l_\ast^2\, \Gamma\left(\, -1\ ; l_\ast^2/
 r_H^2\, \right)\ ,\quad r_H=r_H\left(\, s\,\right)
 \label{crsect}
 \end{equation}
 where, $\Gamma\left(\, -1\ ; l_\ast^2/r_H^2\, \right)$ is the upper incomplete
 Gamma function defined as
 \begin{equation}
 \Gamma\left(\, \alpha\ ; x\, \right)\equiv \int_x^\infty dt\, t^{\alpha-1}
 e^{-t}
 \end{equation}
 $\Gamma\left(\, \alpha\ ; x\, \right)$ is a smooth function with the following
 behavior.  For $s\to 0$ we get
  \begin{equation}
   \sigma\left(\, s\, \right)\approx \pi \, l_\ast^2\,
   \left(\, \frac{r_H}{l_\ast}\, \right)^4\,
   e^{-l_\ast^2/r_H^2}\longrightarrow 0
  \label{lowen}
  \end{equation}
  Equation (\ref{lowen}) means that the production of arbitrary small
  balck holes at low energy is zero, as we expected.\\
  The high-energy limit of (\ref{crsect}) is obtained by means of the
  asymptotic 
  \begin{equation}
  \frac{\Gamma\left(\, \alpha\ ; x\,\right)}{x^\alpha}\to -\frac{1}{\alpha}\ ,\quad
  x\to 0
  \end{equation}
  We thus reproduce the semi-classical black-disk cross section
  \begin{equation}
   \sigma\left(\, s\, \right)\approx \pi \, l_\ast^2\times 
   \left(\, \frac{l_\ast^2}{r_H^2}\,\right)^{-1}=
   \pi \,r_H^2\left(\, s\ ; l_\ast\,\right)
  \end{equation}
   In summary, at energy above the higher dimensional unification
   scale, $\sqrt{s}> M_\ast=l_\ast^{-1}$ the production cross section takes on
   the semi-classical black-disk form, while it drops to zero very
   quickly for $\sqrt{s}< M_\ast$ (see Fig. \ref{fig1}).

   \section{Quantum gravity improved black holes.}
     In the following
   analysis, we take a further step to improve
   $ \sigma\left(\, s\, \right) $ by considering that in the
   presence of a minimal length, whatever it is, the spacetime
   geometry itself is subjected to modifications.
   This step is motivated by the fact that semiclassical black holes offer reliable spacetime descriptions (e.g. in (\ref{rhd})) only when their masses are well above the fundamental mass, while they  become increasingly inaccurate for energies at or just above the fundamental scale.   This particular phenomenology is the goal of the present investigation. 
   To this purpose, we recall that in recent years there have 
   been several attempts to incorporate in black hole spacetimes the presence of 
   quantum gravity effects through effective quantum geometries 
   \cite{NCBHs,LQBHs,ASGBHs,KZ,NLGBHs}. The resulting 
   metrics (QGBHs) tend to agree on some highly desired 
   general characters like the absence of any curvature singularity and a 
   thermodynamically stable cooling down at the end of the evaporation 
   \cite{rob}. In addition one finds that QGBHs do not suffer from a 
   relevant back reaction, a fact that permits a safe employment of 
   quantum field theory in curved space without any breakdown of 
   the formalism. QGBHs have distinctive emission spectra: they tend to 
   emit larger number of softer particles than semi-classical black holes 
   with a suppressed bulk emission \cite{NiW11}.
 As a consequence QGBHs are a natural alternative 
   to improve the scenario at hand.

\begin{figure}
 \begin{center}
\includegraphics[height=10cm]{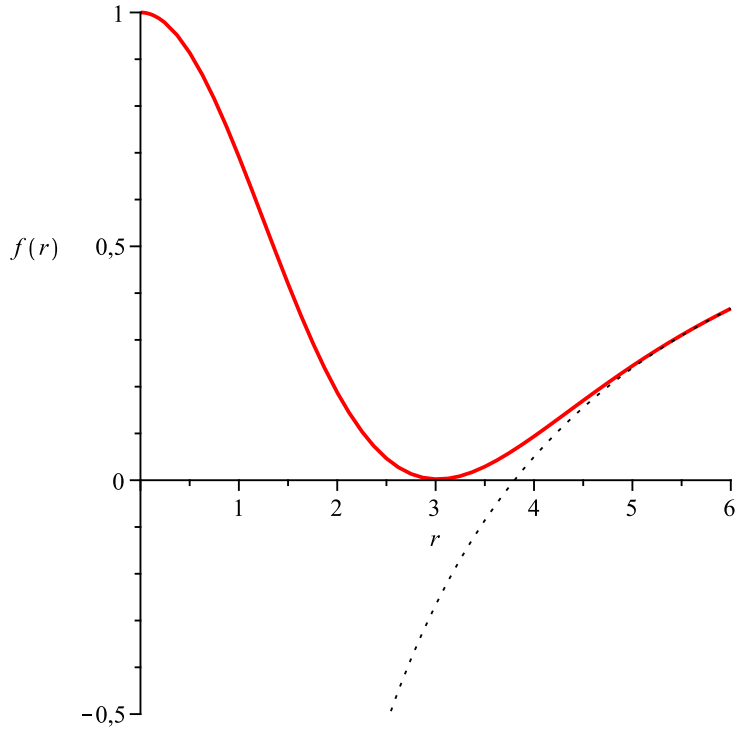}

       \caption{\label{fig1bis} {The function $f(r)$ in $l_\ast$-units. The solid curve is QGBH extremal configuration, while the dotted curve is the corresponding classical black hole having the same mass $\Mbh=M_0$.}}
     \end{center}
  \end{figure}  
   
    A common features of such models is the introduction of a generic
    minimum length $l$,
   which is obtained by means of a modified structure of metric coefficients
     \begin{eqnarray} 
   && ds^2=-f\left(\, r\,\right)\, dt^2 +f^{-1}\left(\, r\,\right)\, dr^2 
   +r^2d\Omega^2 \ , 
   \\ 
   && f\left(\, r\,\right)\equiv 1-{\cal G}_d(r)\left(\frac{2M_{\mathrm{BH}}}{r^{d-2}}\right). 
   \end{eqnarray} 
The function ${\cal G}_d(r)$ models quantum gravity corrections and is subject to the following model independent constraints (for more detailed discussions see \cite{rob}):
\begin{enumerate}[i)]
\item for $r\gg l$ the function ${\cal G}_d(r)$ matches its classical value, i.e., ${\cal G}_d(r)\to G_\ast$; 
\item for $r\sim l$ the function ${\cal G}_d(r)$ enters an ``asymptotically safe regime'' by decreasing with respect its classical value i.e., ${\cal G}_d(r)< G_\ast$ in order to allow the horizon extremization, i.e., $f(r_0)=f^\prime(r_0)=0$ (see Fig. \ref{fig1bis} and (\ref{system}) for more details) and a black hole phase transition to a positive heat capacity cooling down phase;
\item for $r\lesssim l$ the function ${\cal G}_d(r)$ is vanishing in order to improve the curvature singularity, i.e., ${\cal G}_d(r)\sim {\cal O}\left((r/l)^{d-2}\right)$. 
\end{enumerate}

  The minimum length $l$ is not fixed {\it a priori} but is assumed to be
   in the range $ l_{Pl.} \le l \le l_\ast $, where the Planck length $l_{Pl.}$ is 
   the usual four-dimensional gravitational length scale and $l_\ast$ is  
   its higher dimensional counterpart at the TeV scale.
   Such a minimal length bears
   critical importance to the production cross-section of any QGBHs
   that may result in high energy collisions
\cite{Hossenfelder:2003jz}.   
   It would be tempting to say that $\sigma\left(\, s\, \right)$ is given by 
   equation (\ref{crsect}), with $l_\ast$ replaced by $l$. This is almost 
   correct as it properly takes into account the role of $l$, but ignores the existence of 
   a minimum mass $M_0$ below which QGBHs do not form \cite{hep-ph/0109085}. This minimum mass is a 
   common feature of QGBHs and corresponds to the mass of the extremal configuration. The value of $M_0$ can be calculated by solving the system 
\begin{equation}
\left\{ \begin{array}{ll}
f(r)=1-{\cal G}_d(r)\left(\frac{2M_{\mathrm{BH}}}{r^{d-2}} \right)=0 \nonumber \\
f^\prime(r)=\frac{2M_{\mathrm{BH}}}{r^{d-3}}\left(d-2-r\frac{{\cal G}^\prime_d}{{\cal G}_d}\right)=0 \nonumber \\
\end{array}
\right. \\ \label{system}
\end{equation}
in terms of $M_0=M_0(r_0)$.   
   Since
   the creation channels opens up only 
   for $\sqrt{s}\ge M_0$, 
   we thus need to include this threshold condition  
   in the cross section through a second step-function  
   \begin{equation} 
   \sigma\left(\, s\, \right)=\pi\, l^2\, \Gamma\left(\, -1\ ; l^2/r_H^2\, \right) 
     \,  \Theta(\sqrt{s}-M_0)~~.
       \label{sharp} 
   \end{equation} 
   Equation (\ref{sharp}) describes the sharp opening of the production channel 
   at the energy $M_0$, but quantum mechanics introduces uncertainty and 
   makes the step less sharp.  The way the edges of the step are smoothed is 
   determined by the ``golden rule'', derived by previous investigations 
   in noncommutative geometry, leading to regular line elements \cite{NCBHs}. 
   In a nut-shell, the presence of a minimal length (whatever it is its origin) 
   translates into the replacements of Dirac delta functions into minimal width 
   Gaussian distributions. 
   Furthermore, since the Dirac delta is the ``derivative'' of the Heaviside function, 
   it can be shown that in the framework of a minimal length 
   a modified step function can be defined without the limit $l \rightarrow 0$:
 \begin{eqnarray}
\Theta(x)\rightarrow\Theta_l(x)&=&\frac{1}{(4\pi l^2)^{1/2}} \int_{-\infty}^x e^{-y^2/4l^2}dy	 \nonumber \\
   &=& \frac{1}{2}+\frac{1}{2} \mathrm{erf}(x/2l)~~. \nonumber
\end{eqnarray}
To derive the profile of the new cross section, we also need to determine the horizon radius by solving the equation $f(r)=0$. This can best be done by iteration on the expression
  \begin{equation} 
   r^{d-2}_H=2\sqrt{s}\ {\cal G}_d(r_H) ~~,
   \end{equation}
       giving the terms   
 \begin{eqnarray}
   0^{\rm th}~{\rm order}~~~ &\Longrightarrow~~~ &r_{H(0)}=\left(\, 2G_\ast\,\sqrt{s}\,\right)^{\frac{1}{d-2}}
   \nonumber\\ \label{0order}\\
     1^{\rm st}~{\rm order} ~~~ &\Longrightarrow~~~    &r_{H(1)}= r_{H(0)}\,\left[\, 
    \frac{{\cal G}_d(r_{H(0)})} 
    {G_\ast}\,\right]^{\frac{1}{d-2}} 
        \nonumber\\ \label{1order} 
    \end{eqnarray}
   As a first step, we consider just the $0$-th order result (\ref{0order}) 
    and we truncate the iteration process there. 
    For illustrative purposes we can assume $M_0\sim M_\ast.$  In the next section we will show that, though the $0$-th order approximation of the radius may be still acceptable, the assumption of the threshold mass becomes inadequate when quantum gravity corrections of black hole metrics are properly taken into account.

  Using $x=\sqrt{s}-M_0$ and $r_H \simeq (2G_\ast \sqrt{s})^\frac{1}{d-2}$ at
   first order, we obtain  
  \begin{eqnarray}
\frac{\sigma\left(\, s\, \right)}{\pi r_H^2(s)} &=&  \frac{l^2}{(2 G_\ast \sqrt{s})^{\frac{2}{d-2}}} \  
     \, \Gamma\left(\, -1\ ;\frac{l^2}{(2 G_\ast \sqrt{s})^{\frac{2}{d-2}}}\, \right) \nonumber\\ &\times &
      \, \Theta_l(\sqrt{s}-M_0). 
      \label{sfinal}
\end{eqnarray}
This implies $d$-dependent cross-section suppressions, dependent also on the presence 
of a mass threshold \cite{review}.  

   %
    %
   We stress the different roles of the two functions in (\ref{sfinal}). 
    The first ($\Gamma(...)$) comes form the hoop conjecture once the 
    impact parameter is integrated over with a proper short-distance cut-off, while 
    the second ($\Theta_l(...)$) describes the smooth opening of the production 
    channel for  $\sqrt{s}>M_0$.   Generically, these can be understood to 
    represent the outgoing (BH) and incoming 
    (beam) states of the collision, respectively.


   \section{Black hole parameters/energy relations. }

In the previous section we estimated the cross section by considering $0$-th order parameters. This basically corresponds to ignoring the exact nature of quantum corrections to obtain an approximate expression for the cross section which, at the given order, turns out to be model independent.  This procedure can be improved in order to obtain more accurate results. To this purpose, one has to specify a given model of QGBH and determine horizon radii and minimum masses.  In doing so, one opens the possibility of discriminating among the proposed quantum gravity corrections in the class of QGBHs by comparing the resulting cross sections with experimental data. 

%
         %

  To illustrate the procedure, we will focus on noncommutative geometry inspired 
   black holes (NCBHs) only \cite{NCBHs}, leaving the analysis of the whole array
   of QGBHs in future contributions.

   This choice is motivated by the following reasons. NCBHs are for now the 
   richest family of QGBHs, since they include the higher-dimensional
   charged \cite{charged,chargedhigher}, 
   spinning \cite{rotating} and charged-spinning \cite{crotating} 
   solutions. NCBHs are thus the only family of solutions 
   able to describe the complete life cycle of a black hole from its formation to the end of the 
   evaporation, a crucial necessity for phenomenological studies \cite{pheno,pheno2}. Second, 
   NCBHs not only capture the two primary features of QGBHs ({\it i.e.} regularity of the 
   manifold and cooling down phase at the end of the evaporation 
   \cite{review}), but being a subfamily of 
   another class of QGBHs -- namely black holes in nonlocal gravity theories \cite{NLGBHs} -- 
   pave the way to model independent phenomenological conclusions. 
  
   The simplest realization of a noncommutative $(d+1)$-dimensional spacetime due to a
   collapsing parton system is given by equation 
   \begin{eqnarray} 
   && ds^2=-f\left(\, r\,\right)\, dt^2 +f^{-1}\left(\, r\,\right)\, dr^2 
   +r^2d\Omega^2 \ , 
   \label{ncschw}\\ 
   && f\left(\, r\,\right)\equiv \left(1-\frac{2\Mbh }{M_*^{d-1}r^{d-2}}\  
   \frac{\gamma(d/2;r^2/4l^2)}{\Gamma(d/2) }\right) 
   \end{eqnarray} 
   where, to avoid notational confusion, 
   we again indicate with $l$ the minimal length related to the size of 
   spacetime discretization cells, while we keep $G_\ast=M_\ast^{1-d}$ 
   for the gravitational  coupling.
   The above line element can be equivalently identified by the function
    \begin{equation}
{\cal G}_d(r)= \frac{1}{M_*^{d-1}}\  
   \frac{\gamma(d/2;r^2/4l^2)}{\Gamma(d/2) }
   \label{Gnc}
\end{equation}
   in which deviations from the traditional line 
   element are taken into account by the lower incomplete Gamma function 
   \begin{equation} 
   \gamma(d/2\ ; r^2/4l^2)=\int_0^{r^2/4l^2}dt \ t^{d/2-1}e^{-t}~~. 
   \end{equation} 
   While the above metric can exhibit  Killing, Cauchy and degenerate horizons, 
   the singularity at $r=0$ has been removed by ``spreading'' the 
   total mass energy $\Mbh$ over a region of linear size $l$.
   In addition the mass spectrum is bounded from below by an extremal configuration
   which exists even in the case of neutral, non-spinning black holes.  As expected for any QGBH, the ``classical'' relation (\ref{rhd}) 
   between horizon radii and $\sqrt{s}$ is still valid in the high energy limit $\sqrt{s}\gg l^{-1}$.  Conversely, 
   relevant quantum gravity deviations occur for $\sqrt{s}\sim l^{-1}$.
   

    %
    %
    %
%
%
%
    %
    %
    %
    %
%
    %
%

The choice (\ref{Gnc}) for the profile of ${\cal G}_d$ lets us determine the quantities $r_0$ and $M_0$ in terms of $l$  from the system
\begin{equation} 
\left\{ \begin{array}{lll}
    f^\prime\left(\, r_0\,\right)=0~~~&\rightarrow~~~ &  
    r_0=\frac{\left(\, d-2\,\right)^{1/d}}{2^{(1-d)/d}}\, l\ \times
        \nonumber\\  
   && \left[\,\gamma\left(\, \frac{d}{2}\ ; \frac{r^2_0}{4l^2}\,\right)\,\right]^\frac{1}{d}  
    e^{r_0^2/4dl^2} \nonumber\\ 
    f\left(\, r_0\,\right)=0~~~&\rightarrow~~~ & 2G_\ast\, M_0= 
    r_0^{d-2} 
     \frac{\Gamma\left(\, \frac{d}{2} \,\right)}{\gamma\left(\,\frac{d}{2}\ ;\frac{r^2_0}{4l^2}\,\right)} 
     \nonumber\\
       \end{array}
\right. \\ \label{system2}
       \end{equation}
   Note that (\ref{system2}) can be solved only through 
   numerical methods, whose results are given in Table~\ref{ta1} \cite{review}. 

%

      \begin{table}[pht!]
 {\begin{tabular}{@{}ccccccccc@{}} 
 $d$
 &   4 &  5 &  6 &  7 &  8 &  9 \\
 \colrule $M_0$ (TeV)
 & $6.7$  & $24$  & $94$  & $3.8\times 10^2$
 & $1.6\times 10^3$  & $7.3\times 10^3$  \\
\colrule $r_0$ ($10^{-4}$ fm)  &  $2.68$  & $2.51$  & $2.41$  & $2.34$  &
$2.29$  & $2.26$   \\
\end{tabular} 
\caption{$M_0$ and $r_0$ for different values of $d$ and $l=l_\ast=M_\ast^{-1}=1\ \mathrm{TeV}^{-1}$ For $d=10$ one finds $M_0 \simeq 3.4 \times 10^4\ \mathrm{TeV}$ and $r_0\simeq 4.40 \ \mathrm{TeV}^{-1}$.}
\label{ta1}}
\end{table}

      \begin{table}[pht!]
 {\begin{tabular}{@{}ccccccccc@{}} 
 $d$
 &   4 &  5 &  6 &  7 &  8 &  9 \\
 \colrule $M_0$ (TeV)
 & $15.8$  & $102$  & $581$  & $3.02\times 10^3$
 & $1.48\times 10^4$  & $6.91\times 10^4$  \\
\colrule $r_0$ ($10^{-4}$ fm)  &  $2.68$  & $2.51$  & $2.41$  & $2.34$  &
$2.29$  & $2.26$   \\
\end{tabular} 
\caption{$M_0$ and $r_0$ for different values of $d$ and $l=l_\ast=M_\ast^{-1}=1\ \mathrm{TeV}^{-1}$ according to Myers-Perry definition for the fundamental mass $M_\ast$ \cite{mp86}. For $d=10$, $M_0\simeq 3.13\times 10^5\ \mathrm{TeV}$ and $r_0\simeq 2.23 \ \mathrm{TeV}^{-1}$. }
\label{ta1-2}}
\end{table}

      \begin{table}[pht!]
 {\begin{tabular}{@{}cccccccccc@{}} 
 $d$
 & &  4 &  5 &  6 &  7 &  8 &  9 & 10\\
 \colrule $M_0$ (TeV)
& & $63.2$  & $65.2$  & $58.8$  & $48.6$
 & $37.9$  & $28.2$ & $20.3$  \\
\colrule $r_0$ ($10^{-4}$ fm) & &  $2.68$  & $2.51$  & $2.41$  & $2.34$  &
$2.29$  & $2.26$  & $2.23$ \\
\end{tabular} 
\caption{$M_0$ and $r_0$ for different values of $d$ and $l=l_\ast=M_\ast^{-1}=1\ \mathrm{TeV}^{-1}$ using the Particle Data group notation for the fundamental mass $M_\ast$ \cite{pheno2}.}
\label{ta1-3}}
\end{table}

   \begin{figure}
 \begin{center}
    \includegraphics[height=10cm]{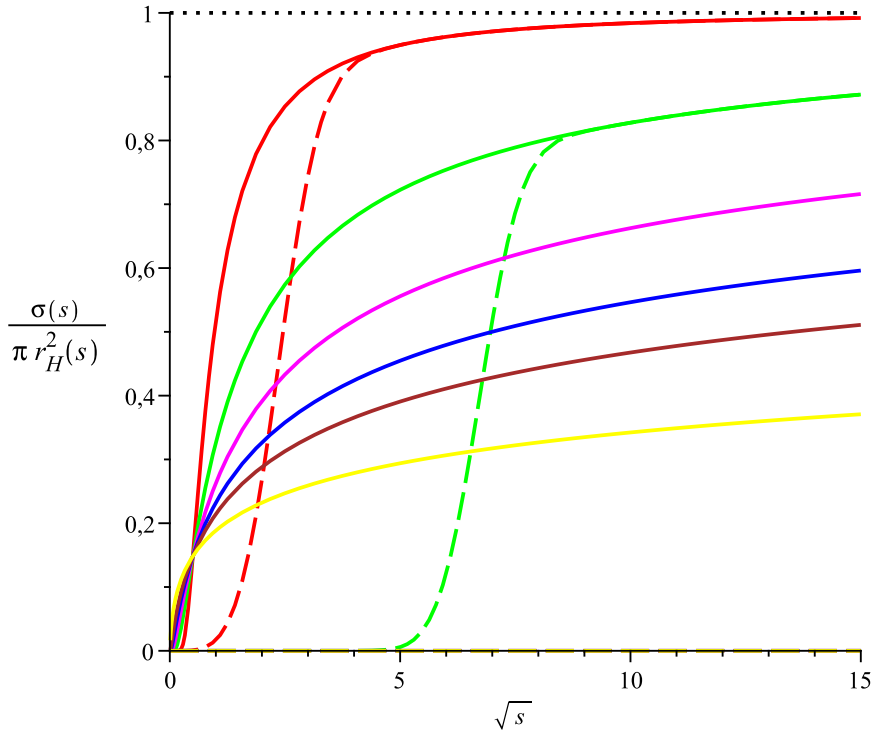}
         \caption{\label{csplot}The ratio of $\sigma(s)$ with and without threshold masses versus the classical value $\pi r^2_H(s)$ as a function of $\sqrt{s}$ in $M_\ast$-units ($M_\ast=l^{-1}=10^{16} \ \mathrm{TeV}$ for $d=3$ and $M_\ast=l^{-1}=1\ \mathrm{TeV}$ for $d>3$). From top to bottom (solid): $d=3,\ 4,\ 5, \ 6, \ 7$ and $10$ without threshold masses $M_0$. Dashed curves take into account the the threshold masses $M_0$. The dotted curve is the classical black disk cross section which has the same profile for any $s$. }
     \end{center}
  \end{figure}  

The discrepancy between $M_0$ and $M_\ast$ underlines the need to properly account for black hole quantum gravity corrections beyond the rudimentary assumption $M_0\sim M_\ast$.

We stress the above values depend on the definitions of the fundamental scale $M_\ast$, which may differ for multiplicative 
 constants (see \cite{NiW11} for more detailed comments on the interrelationship between them). In Table~\ref{ta1-2} and \ref{ta1-3}  we show the black hole threshold parameters according to two other major definitions of the fundamental mass.  For the cases in Table~\ref{ta1} and \ref{ta1-2}, 
the minimum mass increases with spatial dimensionality,  which at LHC energies would yield a virtually vanishing cross-section for $d\geq 6$.
Curiously, according to the Particle Data Group notation (Table~\ref{ta1-3}), we find the minimum mass decreases with spatial dimensionality for any $d\geq 5$. When $d=10$ we find the most promising case, corresponding to a minimum mass of roughly $6~$TeV above the  maximum LHC centre-of-mass energy.


One may wonder what is the production rate related to the final formula
 \begin{equation} 
   \sigma\left(\, s\, \right)=\pi\, l^2\, \Gamma\left(\, -1\ ; l^2/r_H^2(s)\, \right) 
     \,  \Theta_l(\sqrt{s}-M_0)~~.
       \label{smoothcross} 
   \end{equation} 
 assuming the LHC's current peak luminosity. Unfortunately, we cannot determine the horizon radius $r_H$ as a function of $\sqrt{s}$ in a closed form. From the data presented in the above tables, however, we see that quantum gravity deviations are largely a function of the threshold energy.   Conversely, the horizon radii are less sensitive to non-classical effects and approach the range of classical values $\sim 10^{-4}~$fm even in the  realm of maximum corrections, {\it i.e.} in the vicinity of the extremal configuration.   This is the case irrespective of the definition of the fundamental mass, since $r_0$ is determined through the first equation of the system (\ref{system2}). 
 
\begin{figure}
 \begin{center}
    \includegraphics[height=10cm]{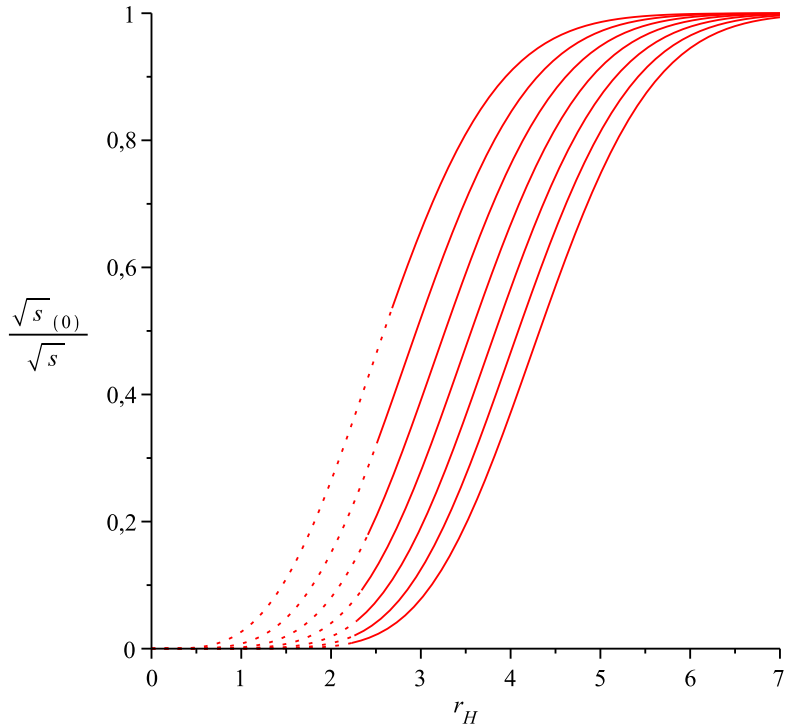}
         \caption{\label{beamen} The ratio $\sqrt{s}_{(0)}/\sqrt{s}$ of the parton energies required for the formation of classical ($0$-th order) and non-classical horizons, in units $l=1$. Curves from botton to top refer to $d=3-10$.   All ratios are less than unity, indicating quantum gravity effects slow down production rates by requiring more energy for horizon formation.  When $r_H\geq 6l$ quantum gravity corrections quickly die off, while for smaller radii the dotted curves indicate the regime where non-classical horizon do not form for the presence of threshold energies.}
     \end{center}
  \end{figure}

 We can show this from the equation $f(r_H)=0$, by considering the parton energy to contribute to both classical ($0$-th order) and non-classical horizons, \textit{i.e.} $\sqrt{s}_{(0)}$ and $\sqrt{s}$ respectively, whose ratio 
\begin{equation}
\frac{\sqrt{s}_{(0)}}{\sqrt{s}}=\frac{\gamma(d/2; r_H^2/4l^2)}{\Gamma(d/2)}
\end{equation} 
is plotted in Figure~\ref{beamen}.  We see that the above approximation works very well not only in the high energy regime (\textit{i.e.} for $r_H\geq 6l$), but also below the production threshold where the the function $\Theta_l(\dots)$  excludes the discrepancies arising from the $0$-th order approximation of the actual non-classical horizon.  
It is therefore not  difficult to improve the result in (\ref{sfinal}) by considering the correct threshold masses, while keeping horizon radii approximated at the $0$-th order. 
          \begin{equation}
 r_H\sim r_{H(0)}=\left(\, 2G_\ast\,\sqrt{s}\,\right)^{\frac{1}{d-2}}
 \end{equation}

    \begin{table}[pht!]
 {\begin{tabular}{@{}c||lc|c|c|c|c@{}} 
 $\sqrt{s}$= &  & $90 \ \mathrm{TeV}$ &  $91 \ \mathrm{TeV}$ &  $92 \ \mathrm{TeV}$ &  $94 \ \mathrm{TeV}$ & $100 \ \mathrm{TeV}$  \\
 \colrule $d=5$& & $129\ \mathrm{s}^{-1}$  & $130\ \mathrm{s}^{-1}$  & $131\ \mathrm{s}^{-1}$  & $132\ \mathrm{s}^{-1}$
 & $138\ \mathrm{s}^{-1}$ \\  
  \colrule $d=6$& & $12\ \mathrm{yr}^{-1}$  & $0.55\ \mathrm{h}^{-1}$  & $6\ \mathrm{min}^{-1}$  & $23\ \mathrm{s}^{-1}$
 & $51\ \mathrm{s}^{-1}$  \\ 
 \colrule $d=7$& & $< 1\ \mathrm{T_U}^{-1}$  & $< 1\ \mathrm{T_U}^{-1}$  & $< 1\ \mathrm{T_U}^{-1}$  & $< 1\ \mathrm{T_U}^{-1}$
 & $< 1\ \mathrm{T_U}^{-1}$   
\end{tabular} 
\caption{The number of black holes per unit of time $\dot{N}$ for different values of $d$ and $\sqrt{s}$.  Here, $\mathrm{T_U}=13.7 \ \mathrm{Gyr}$ is the age of the universe. Values of $\dot{N}$ have been calculated from (\ref{sfinal}) by considering threshold masses as in Table~\ref{ta1}, the fundamental mass  $M_\ast= l^{-1}=l_\ast^{-1}=1\ \mathrm{TeV}$, the current LHC luminosity $L\sim 3.65 \times 10^{38}\ \mathrm{m}^{-2}\mathrm{s}^{-1}$ and the classical black disk cross section for each $d$ and value of energy $\sqrt{s}$.} 
\label{ta2}}
\end{table}

In the process, the production rate is mildly overestimated by virtue of the fact ${\cal G}_d\leq G_\ast$ implies  $r_{H(0)}\gtrsim r_H$ (or correspondingly $\sqrt{s}_{(0)}<\sqrt{s}$), \textit{i.e.}, classical horizon formation requires less energy).  
Consequently, use of (\ref{sfinal}) is justified under proper choice of $M_0$.
Tighter constraints may be obtained by simply proceeding with a more sophisticated approximation for the horizon radius, \textit{i.e.}, $r_{H(1)}$. 
Figure~\ref{csplot} demonstrates the resulting profile of cross sections for the choice of fundamental mass $M_\ast$ of Table~\ref{ta1}. 
In light of the resulting rates being extremely suppressed at LHC energies for all $d$, 
 we can make just an example of a hypothetical collision at energies $\sqrt{s}=90-100\ \mathrm{TeV}$. 
In Table~\ref{ta2}, we show the black hole production rate $\dot{N}$ for varying energy $\sqrt{s}$ and number of dimensions $d$.  Note the data are very sensitive to both $\sqrt{s}$ and $d$.   For $d=5$, energies above the production threshold $\sim 24\ \mathrm{TeV}$ and $\dot{N}$ saturate at the black disk result.      Remarkably for $d=7$ and energies below the production threshold $\sim 380\ \mathrm{TeV}$,  the rate $\dot{N}$ is so low that the production time for a single black hole would be greater than the present age of the universe ({\it i.e.} $\rm T_U \sim 13.7 \ \mathrm{Gyr}$). Finally for $d=6$, we are at energies close to the production threshold $\sim 94\ \mathrm{TeV}$ and $\dot{N}$ varies dramatically with $\sqrt{s}$ ranging from formation time scale of a second to a month.  This example shows how in principle the black hole production described by (\ref{sfinal}) can also be used to indirectly determine the number of dimensions $d$.

As a check of the huge variation of $\dot{N}$, we can write the near-threshold cross section for the production of extremal black holes. For $\sqrt{s}\sim M_0$, the leading term reads 
     \begin{eqnarray}
\sigma\left(\, s\, \right) \sim  \pi l^2  
     \, \Gamma\left(\, -1\ ;l^2/r_0^2 \, \right) 
      \, \left[\frac{1}{2}+\frac{1}{2l}(\sqrt{s}-M_0)\right]~~, 
      \label{sfinal_lead}
\end{eqnarray}
which describes the approximately linear behavior of the cross section near the production threshold.   The above formula can be employed to improve the results of Table~\ref{ta2} when considering rates at near threshold energies, \textit{i.e.}, for $d=6$ and $\sqrt{s}=90-100\ \mathrm{TeV}$. The corresponding values in Table~\ref{ta2-2} show that the $0$-th order approximation can capture the orders of magnitude of productions rates even in  this limit.   As expected, the quantum gravity corrections of horizon radii result in slightly suppressed rates.
%
%

    \begin{table}[pht!]
 {\begin{tabular}{@{}c||lc|c|c|c|c@{}} 
 $\sqrt{s}$= &  & $90 \ \mathrm{TeV}$ &  $91 \ \mathrm{TeV}$ &  $92 \ \mathrm{TeV}$ &  $94 \ \mathrm{TeV}$ & $100 \ \mathrm{TeV}$  \\
  \colrule $d=6$& & $3.7\ \mathrm{yr}^{-1}$  & $0.20\ \mathrm{h}^{-1}$  & $2.2\ \mathrm{min}^{-1}$  & $8\ \mathrm{s}^{-1}$
 & $16\ \mathrm{s}^{-1}$  \\ 
\end{tabular} 
\caption{The number of black holes per unit of time $\dot{N}$ for $d=6$ and $\sqrt{s}=90-100\ \mathrm{TeV}$.   Values of $\dot{N}$ have been calculated from (\ref{sfinal_lead}) by considering threshold mass $M_0=94\ \mathrm{TeV}$ and the extremal black hole radius $r_0=2.41\ \mathrm{TeV}^{-1}$ as in Table~\ref{ta1}, the fundamental mass $M_\ast= l^{-1}=l_\ast^{-1}=1\ \mathrm{TeV}$ and the current LHC luminosity $L\sim 3.65 \times 10^{38}\ \mathrm{m}^{-2}\mathrm{s}^{-1}$.} 
\label{ta2-2}}
\end{table}

Estimates of $\dot{N}$ for the Myers-Perry definition will not give higher production rates due to the heavier threshold masses, as is evident from Table~\ref{ta1-2}. Consequently, it may be interesting to explore the case of the Particle Data Group definition, whose gravitational coupling constant turns out to be
\begin{equation}
G_\ast\to G_\ast=\frac{2^{d-4}}{d-1}\  \pi^{(d-6)/2} \Gamma(d/2) M_\ast^{1-d}~~.
\end{equation}

Table~\ref{ta3} lists the production rates for the case $d=10$. Despite the low threshold mass, we find that at typical LHC energies the production of black hole turns out to be improbable: at $14\ \mathrm{TeV}$, roughly one black hole every $60$ million years would be produced in particle detectors. Due to the vicinity to the threshold mass, however, the data strongly vary and already at $16\ \mathrm{TeV}$ one finds a promising value of one black hole per month being produced.   The above results can be improved by using (\ref{sfinal_lead}), which works better near threshold and provides tighter constraints.    In this case the maximum LHC beam energy significantly lowers the production rate to one black hole every $225$ million years, while at $16\ \mathrm{TeV}$ one finds a rate of approximately one black hole every 4  months.  Table~\ref{ta3-2} shows the complete results that confirm how the $0$-th order approximation can be considered valid to estimate orders of magnitude.

   \begin{table}[pht!]
 {\begin{tabular}{@{}c||lc|c|c|c|c@{}} 
 $\sqrt{s}$= &  & $10 \ \mathrm{TeV}$ &  $14 \ \mathrm{TeV}$ &  $16 \ \mathrm{TeV}$ &  $17 \ \mathrm{TeV}$  & $20 \ \mathrm{TeV}$  \\
 \colrule $d=10$& & $< 1\ \mathrm{T_U}^{-1}$  & $0.016\ \mathrm{Myr}^{-1}$ & $13\ \mathrm{yr}^{-1}$  & $2.3\ \mathrm{h}^{-1}$  & $30\ \mathrm{s}^{-1}$  
\end{tabular} 
\caption{The number of black holes per unit of time $\dot{N}$ as a function of $\sqrt{s}$ for $d=10$, calculated from  (\ref{sfinal}) by considering threshold mass $M_0=20.3\ \mathrm{TeV}$ as in Table~\ref{ta1-3} (Particle Data Group notation), the fundamental mass  $M_\ast=l^{-1}=l_\ast^{-1}=1\ \mathrm{TeV}$, the current LHC luminosity $L\sim 3.65 \times 10^{38}\ \mathrm{m}^{-2}\mathrm{s}^{-1}$ and the classical black disk cross section for each value of energy $\sqrt{s}$. }
\label{ta3}}
\end{table}

    \begin{table}[pht!]
 {\begin{tabular}{@{}c||lc|c|c|c|c@{}} 
 $\sqrt{s}$= &  & $10 \ \mathrm{TeV}$ &  $14 \ \mathrm{TeV}$ &  $16 \ \mathrm{TeV}$ &  $17 \ \mathrm{TeV}$  & $20 \ \mathrm{TeV}$  \\
 \colrule $d=10$& & $< 1\ \mathrm{T_U}^{-1}$  & $0.004\ \mathrm{Myr}^{-1}$ & $3.2\ \mathrm{yr}^{-1}$  & $0.52\ \mathrm{h}^{-1}$  & $6.6\ \mathrm{s}^{-1}$  
\end{tabular} 
\caption{The number of black holes per unit of time $\dot{N}$ as a function of $\sqrt{s}$ for $d=10$, calculated from  (\ref{sfinal_lead}) by considering threshold mass $M_0=20.3\ \mathrm{TeV}$ and the extremal horizon radius $r_0=2.23\ \mathrm{TeV}$ as in Tab. \ref{ta1-3} (Particle Data Group notation), 
the fundamental mass $M_\ast= l^{-1}=l_\ast^{-1}=1\ \mathrm{TeV}$ and the current LHC luminosity $L\sim 3.65 \times 10^{38}\ \mathrm{m}^{-2}\mathrm{s}^{-1}$. }
\label{ta3-2}}
\end{table}

\section{Final remarks.}
We have presented a first step in modeling black hole production in a 
post-semiclassical limit, with quantum gravity effects being introduced by 
a minimal length $l$.  Black production cannot occur among the variety of quantum gravity corrections we have considered, implemented below their respective threshold masses.  We have provided a complete analysis of the associated cross-sections for the case of NCBHs.  The related black hole production rates are highly sensitive to the value of the threshold masses, which vary non only according to the number of extra dimensions but also to the definition of the fundamental mass. Our results show that that microscopic 
black hole production is not a likely scenario for energies below 100~TeV 
with a minimum $d=6$ spatial dimensions. However for the case of Particle Data Group definition of the fundamental mass, we find that the LHC would be just a couple of TeV below a reasonable production rate, provided that $d=10$. Our approach assumes the extra-dimensional characteristics of spacetime are those of the ADD mechanism \cite{add}, but we acknowledge that other terascale gravity
models also produce similiar phenomenology, including Randall-Sundrum \cite{rizzo},
ungravity \cite{jrm1}, {\it etc.}. 
Additionally, even if we believe to have found the correct method to study these issues, our conclusions cannot be considered definitive: we still miss a complete analysis of all the remaining quantum gravity corrected black holes whose threshold masses might be at the reach of the LHC.  Consequently, our result can be used by reversing the logic: instead of predicting production rates, one may determine the correct quantum gravity theory from experiment, \textit{i.e.} through the value of the observed threshold mass for non-negligible production rates at LHC.

Whatever the case, even in the most pessimistic scenario such novel phenomenology is 
still potentially observable in ultra-high energy cosmic ray collisions.

\begin{acknowledgements}
JM and PN would like to thank the Dipartimento di Fisica, 
Universit\`a di Trieste
and INFN for their kind hospitality during the initial period of work on 
this project. This work is supported in part by the Helmholtz International 
Center for FAIR within the
framework of the LOEWE program (Landesoffensive zur Entwicklung 
Wissenschaftlich-\"{O}konomischer Exzellenz) launched by the State of 
Hesse (PN), by the European Cooperation in Science and Technology (COST) action MP0905 ``Black Holes in a Violent Universe'' (PN) and also by a Continuing Faculty Grant from Loyola Marymount 
University (JM). 
\end{acknowledgements}

%
%


 \end{document}